# f²IMU-R: Pedestrian Navigation by Low-cost Foot-Mounted Dual IMUs and Inter-foot Ranging

Maoran Zhu, Yuanxin Wu, *Senior Member*, *IEEE,* and Shitu Luo

*Abstract*—**Foot-mounted inertial sensors become popular in many indoor or GPS-denied applications, including but not limited to medical monitoring, gait analysis, soldier and first responder positioning. However, the foot-mounted inertial navigation relies largely on the aid of Zero Velocity Update (ZUPT) and has encountered inherent problems such as heading drift. This paper implements a pedestrian navigation system based on dual foot-mounted low-cost inertial measurement units (IMU) and inter-foot ultrasonic ranging. The observability analysis of the system is performed to investigate the roles of the ZUPT measurement and the foot-to-foot ranging measurement in improving the state estimability. A Kalman-based estimation algorithm is mechanized in the Earth frame, rather than in the common local-level frame, which is found to be effective in depressing the linearization error in Kalman filtering. An ellipsoid constraint in the Earth frame is also proposed to further restrict the height drift. Simulation and real field experiments show that the proposed method has better robustness and positioning accuracy (about 0.1-0.2% travelled distance) than the traditional pedestrian navigation schemes do.**

## I. INTRODUCTION

The pedestrian navigation system (PNS) based on inertial measurement Units (IMU) in GPS-denied environment has abundant applications, including but not limited to medical monitoring and recovery [1], human motion capture for animation [2], industrial inspections, and first responder search and rescue [3]. The IMU-based PNS can generally be classified into the non-foot-mounted type and the foot-mounted IMU type [3, 4]. Generally, the non-foot-mounted PNS detects the foot steps, estimates the heading angle and the step length, and then calculate the position change. Although this method allows IMU to be carried at any part of the body, it relies on an ideal walking model and does not utilize the translational information provided by IMU [5]. Recently, the deep learning based PNS were also proposed for estimating the step length and heading angle [6, 7]. While, the foot-mounted PNS tracks real-time position and orientation by integrating gyroscope and accelerometer measurements. However, an open-loop or free inertial navigation system (INS) will lead to cumulative error over time, especially for low-cost Micro Electro-Mechanical sensors (MEMS).

In order to suppress the cumulative INS error, the most widely used technique is the Zero Velocity Update (ZUPT) method [8], which assumes that the foot-mounted IMU is stationary during the mid-stance phase of walking. Once the mid-stance phase is detected [9, 10], the zero velocity is treated as a pseudo measurement to restrain the INS by a Kalman filter. The ZUPT method can mitigate most of the cumulative error, but the heading drift, as well as the yaw gyro bias, is one of major obstacles for autonomous pedestrian navigation over extended period using the foot-mounted IMU [8, 11, 12].

A number of methods have been proposed to solve this problem. These methods can be divided into the constraint-based methods [13-19] and extra sensor-assisted methods [20-23]. The works presented in [13, 16] attempt to correct the yaw gyro bias by assuming a constant heading if a person walks on a straight line. When the person walks back to the starting position, a zero position update is used in [15]. Some authors utilize an inequality constraint in the Kalman filter in view of the fact that there exits an upper bound on the maximum spatial separation between two feet in a system of dual foot-mounted IMUs [18, 24]. Similarly in [17], the minimum inter-foot distance is assumed constant, and a soft equality constraint is applied in the dual foot-mounted system.

Among the sensor-assisted methods, it is quite common to acquire the heading by using a magnetometer [13, 21]. The work in [20] proposes a PNS system aided by the signal strength of several active RFID tags placed at known locations in a building. In [22], the relative position and attitude between two feet are measured by a foot-mounted camera and an infrared LED and then used to aid the dual foot-mount IMUs. A visual-inertial approach is presented in [23] that fuses the data from the dual foot-mounted IMUs and a head-mounted IMU-camera pair by a batch least-squares algorithm.

An extended abstract was presented at the 57th symposium on Inertial Sensors and Systems (ISS), Braunschweig, Germany, 2020.

This work was supported in part by National Key R&D Program of China (2018YFB1305103) and National Natural Science Foundation of China (61673263).

M. Zhu and Y. Wu are with Shanghai Key Laboratory of Navigation and Location-based Services, School of Electronic Information and Electrical Engineering, Shanghai Jiao Tong University, Shanghai, China, 200240, (e-mail: zhumaoran@sjtu.edu.cn; yuanx_wu@hotmail.com;); S. Luo is with Shenzhen SuperNavi Technology CO., Ltd., Shenzhen, China, 518218 (e-mail: luoshitu@163.com).

| Year | Leading Author | Sensors Used | Major Technical Feature |
|------|----------------|--------------|-------------------------|
| 2011 | Laverne [9] | IMU & Sonar | Inter-foot distance measured by sonar; Tactical IMUs |
| 2012 | Skog [16] | IMU | Upper bound of feet distance imposed by inequality-constraint KF |
| 2013 | Hung [20] | IMU & Camera | Relative feet pose estimated by foot-mounted camera |
| 2015 | Weenk [23] | IMU & Sonar | Inter-foot ranging for gait parameter analysis |
| 2018 | Ahmed [21] | IMU & Camera | Google glass and two foot-mounted IMUs used for body pose estimation by batch-LS method |
| 2019 | Niu [15] | IMU | Minimum constant inter-foot distance imposed by equality-constraint KF |
| 2019 | Wang [24] | IMU & Sonar | Two pairs of sonars for directional inter-foot ranging |
| 2019 | Zhao [17] | IMU | Inverted pendulum gait model |

In [25], a dual foot-mounted IMUs system with the foot-to-foot ranging measurement is presented to estimate the relative foot position for medical monitoring and recovery applications. Despite the authors assertion that their system cannot be used for navigation purposes, an example in [11] has demonstrated that dual foot-mounted tactical-grade IMUs with foot-to-foot ranging can indeed be used for PNS. The directional distance between two feet is employed to aid the PNS system in [26], but the positioning accuracy appears not quite satisfying. For easy reference, Table 1 summarizes current dual foot-mounted autonomous PNS systems and their technical features.

In this paper, we implement a PNS system for autonomous navigation with dual foot-mounted consumer-grade IMUs and ultrasonic inter-foot ranging (named f²IMU-R hereafter) and demonstrate its navigation capability for potential applications. Compared against previous works, the major technical contribution rests on the global observability analysis of the ZUPT and the inter-foot ranging, the Earth-frame (e-frame) mechanization for mitigating the EKF linearization errors and an ellipsoid constraint for suppressing the height drift.

The rest of this paper is organized as follows. Section II reviews the ZUPT-based PNS algorithm in the e-frame. Section III describes the f²IMU-R with the ellipsoid constraint. Global observability analysis and observability improvement due to the inter-foot ranging will be discussed. Sections IV-V report the simulation and the field test results. The conclusion is drawn in Section VI.

## II. INS MECHANIZATION AND ZUPT-AIDED PNS

### A. INS Mechanization

The dynamic equations for a strapdown INS in the e-frame are given by [27]

$$\dot{\mathbf{p}}^e = \mathbf{v}^e \tag{1}$$

$$\dot{\mathbf{v}}^e = \mathbf{C}_b^e \left( \mathbf{f}^b - \mathbf{b}_a - \mathbf{n}_a \right) - 2\boldsymbol{\omega}_{ie}^e \times \mathbf{v}^e + \mathbf{g}^e \tag{2}$$

$$\dot{\mathbf{C}}_b^e = \mathbf{C}_b^e \left( \boldsymbol{\omega}_{eb}^b \times \right) \tag{3}$$

$$\boldsymbol{\omega}_{eb}^b = \boldsymbol{\omega}_{ib}^b - \mathbf{b}_g - \mathbf{C}_e^b \boldsymbol{\omega}_{ie}^e - \mathbf{n}_g, \tag{4}$$

where $\mathbf{p}^e = \begin{bmatrix} x^e & y^e & z^e \end{bmatrix}^T$ and $\mathbf{v}^e$ denote the position and the ground velocity in the e-frame, $\mathbf{C}_b^e$ denotes the body's attitude rotation matrix with respect to the e-frame, $\mathbf{f}^b$ is the specific force measured by accelerometers and expressed in the body frame (b-frame), $\mathbf{b}_a$ is the accelerometer bias, $\boldsymbol{\omega}_{ie}^e = \begin{bmatrix} 0 & 0 & \Omega \end{bmatrix}^T$ is the Earth's rotation rate expressed in the e-frame, $\mathbf{g}^e$ is the gravity vector in the e-frame, $\boldsymbol{\omega}_{eb}^b$ is the body's angular velocity with respect to the e-frame and expressed in the b-frame, the skew symmetric matrix $(\cdot \times)$ is defined so that the cross product $x \times y = (x \times) y$ is satisfied for arbitrary two vectors, $\boldsymbol{\omega}_{ib}^b$ represents the angular velocity measured by gyroscopes and $\mathbf{b}_g$ represents the gyroscope bias. $\mathbf{n}_a$ and $\mathbf{n}_g$ are i.i.d zero-mean Gaussian noises with covariance $\sigma_a^2 \mathbf{I}_3$ and $\sigma_g^2 \mathbf{I}_3$, respectively. And, $\mathbf{I}_n$ denotes a n-dimensional identity matrix. The accelerometer bias $\mathbf{b}_a$ and the gyroscope bias $\mathbf{b}_g$ are modeled as random walks

$$\dot{\mathbf{b}}_g = \mathbf{n}_{bg} \tag{5}$$

$$\dot{\mathbf{b}}_a = \mathbf{n}_{ba} \tag{6}$$

where $\mathbf{n}_{bg}$ and $\mathbf{n}_{ba}$ are i.i.d zero-mean Gaussian noises with covariance $\sigma_{ba}^2 \mathbf{I}_3$ and $\sigma_{bg}^2 \mathbf{I}_3$. The attitude, velocity and position in the e-frame are calculated by integrating Eqs. (1)-(3), following the standard mechanization in the inertial navigation community [27]. In addition, the position in the navigation frame (n-frame), designated as North-Up-East without the loss of generality, can be derived from the e-frame position $\mathbf{p}^e$ by an iterative calculation (e.g. [27] and [28]). And, the mutual transformation of attitude and velocity between the e-frame and the n-frame is straightforward, that is,

$$\mathbf{C}_n^b = \mathbf{C}_e^b \mathbf{C}_n^e, \quad \mathbf{v}^n = \mathbf{C}_e^n \mathbf{v}^e \tag{7}$$

where $\mathbf{C}_e^n$ is the rotation matrix from the navigation frame to the earth frame.

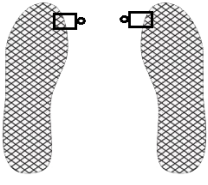

Fig. 1. Pedestrian navigation system with dual foot-mounted IMUs and inter-foot ranging (circle: ultrasonic ranging modules; square: IMU modules).

### B. PNS in State-space Form

The ZUPT-aided PNS assumes that the foot undergoes a stationary phase during every step of walk. A number of static detectors have been proposed based on accelerometer/gyroscope measurements [9, 10]. This paper uses the angular rate energy detector (ARE) to identify the stance phase [9]. Once the stance phase is identified, the zero velocity update can be fused with the above INS mechanization by an error-state Kalman filter.

Define the error state as the state estimate subtracting the true state, i.e. $\delta\mathbf{x} = \hat{\mathbf{x}} - \mathbf{x}$. Specifically, the attitude estimate is defined as being related to the true attitude and the corresponding attitude error $\delta\boldsymbol{\psi}^e$ by $\tilde{\mathbf{C}}_e^b \approx \mathbf{C}_e^b \left( \mathbf{I}_3 + \delta\boldsymbol{\psi}^e \times \right)$,

The error states are collectively defined as

$$\delta\mathbf{x} = \begin{bmatrix} \delta\boldsymbol{\psi}^{eT} & \delta\mathbf{v}^{eT} & \delta\mathbf{p}^{eT} & \mathbf{b}_g^{\ T} & \mathbf{b}_a^{\ T} \end{bmatrix}^T, \qquad (8)$$

Derivation of the first-order error state equation of the process and measurement can be found in [27] and is summarized below. The PNS system model in the state-space matrix form is given by

$$\begin{cases} \delta\dot{\mathbf{x}} = \mathbf{f}\delta\mathbf{x} + \mathbf{dw} \\ \delta\mathbf{v}_v = \mathbf{H}_v \delta\mathbf{x} + \mathbf{n}_v \end{cases} \qquad (9)$$

where the dynamic noise $\mathbf{w} = \begin{bmatrix} \mathbf{n}_g^T & \mathbf{n}_a^T & \mathbf{n}_{bg}^T & \mathbf{n}_{ba}^T \end{bmatrix}^T$ and $\mathbf{n}_v$ denotes the measurement noise with covariance $\sigma_v^2 \mathbf{I}_3$. The matrices involved are defined explicitly as

$$\mathbf{f} = \begin{bmatrix} -\boldsymbol{\omega}_{ie}^e \times & \mathbf{0}_3 & \mathbf{0}_3 & -\mathbf{C}_b^e & \mathbf{0}_3 \\ \left( \mathbf{C}_b^e \mathbf{f}^b \right) \times & -2\boldsymbol{\omega}_{ie}^e \times & \mathbf{0}_3 & \mathbf{0}_3 & \mathbf{C}_b^e \\ \mathbf{0}_3 & \mathbf{I}_3 & \mathbf{0}_3 & \mathbf{0}_3 & \mathbf{0}_3 \\ \mathbf{0}_{6\times3} & \mathbf{0}_{6\times3} & \mathbf{0}_{6\times3} & \mathbf{0}_{6\times3} & \mathbf{0}_{6\times3} \end{bmatrix} \quad \mathbf{d} = \begin{bmatrix} -\mathbf{C}_b^e & \mathbf{0}_3 & \mathbf{0}_3 & \mathbf{0}_3 \\ \mathbf{0}_3 & \mathbf{C}_b^e & \mathbf{0}_3 & \mathbf{0}_3 \\ \mathbf{0}_3 & \mathbf{0}_3 & \mathbf{0}_3 & \mathbf{0}_3 \\ \mathbf{0}_3 & \mathbf{0}_3 & \mathbf{I}_3 & \mathbf{0}_3 \\ \mathbf{0}_3 & \mathbf{0}_3 & \mathbf{0}_3 & \mathbf{I}_3 \end{bmatrix}$$

$$\mathbf{H}_v = \begin{bmatrix} \mathbf{0}_3 & \mathbf{I}_3 & \mathbf{0}_3 & \mathbf{0}_3 & \mathbf{0}_3 \end{bmatrix} \qquad (10)$$

### III. Pedestrian Navigation System with Dual foot-mounted IMUs and Inter-foot Ranging

#### A. PNS Algorithm for Dual Foot-mounted IMUs and Inter-foot Ranging

For a dual foot-mounted PNS, two subsystems, an IMU and an ultrasonic ranging combo module, are attached to the pedestrian feet as shown in Fig. 1. The inter-foot distance is related to the two feet positions by

$$d = \left\| \mathbf{p}_L^e + \mathbf{C}_{b,L}^e \mathbf{l}_L^b - \mathbf{p}_R^e - \mathbf{C}_{b,R}^e \mathbf{l}_R^b \right\| + n_d \triangleq \left\| \Delta\mathbf{l} \right\| + n_d \qquad (11)$$

where the subscripts $L$ and $R$ denote the left foot and the right foot, respectively. $\|\bullet\|$ denotes the vector magnitude, $\mathbf{l}^b$ is the lever arm between the IMU and the ultrasonic unit on the same foot, $n_d$ denotes the distance measurement noise with covariance $\sigma_d^2$. Note that the measurement equation (11) is much simpler than that expressed in the n-frame, which, as shown in later section, is beneficial to reduce the EKF linearization error as well as the complexity of observability analysis. Correspondingly, the dual-IMU joint state vector is defined as

$$\delta\mathbf{X} = \begin{bmatrix} \delta\mathbf{x}_L^T & \delta\mathbf{x}_R^T \end{bmatrix}^T \qquad (12)$$

The dynamic equation is expressed as

$$\begin{cases} \delta\dot{\mathbf{X}} = \mathbf{F}\delta\mathbf{X} + \mathbf{D}\mathbf{W} \\ \delta\mathbf{y} \equiv \begin{bmatrix} \delta\mathbf{y}_{v,L} \\ \delta\mathbf{y}_{v,R} \\ \delta d \end{bmatrix} = \begin{bmatrix} \mathbf{H}_v & \mathbf{0}_{3\times15} \\ \mathbf{0}_{3\times15} & \mathbf{H}_v \\ \mathbf{H}_{d,L} & \mathbf{H}_{d,R} \end{bmatrix} \delta\mathbf{X} + \begin{bmatrix} \mathbf{n}_{v,L} \\ \mathbf{n}_{v,R} \\ n_d \end{bmatrix} \end{cases} \qquad (13)$$

where the dynamic noise $\mathbf{W} = \begin{bmatrix} \mathbf{w}_L^T & \mathbf{w}_R^T \end{bmatrix}^T$ and the matrices involved are defined as

$$\mathbf{H}_{d,L} = \begin{bmatrix} \Delta\mathbf{l}^T \left( \mathbf{C}_{b,L}^e \mathbf{l}_L^b \right) \times & \mathbf{0}_{1\times3} & \Delta\mathbf{l}^T & \mathbf{0}_{1\times6} \end{bmatrix} \Big/ \left\| \Delta\mathbf{l} \right\|$$

$$\mathbf{H}_{d,R} = \begin{bmatrix} -\Delta\mathbf{l}^T \left( \mathbf{C}_{b,R}^e \mathbf{l}_R^b \right) \times & \mathbf{0}_{1\times3} & -\Delta\mathbf{l}^T & \mathbf{0}_{1\times6} \end{bmatrix} \Big/ \left\| \Delta\mathbf{l} \right\| \qquad (14)$$

$$\mathbf{F} = \begin{bmatrix} \mathbf{f}_L & \mathbf{0}_{15} \\ \mathbf{0}_{15} & \mathbf{f}_R \end{bmatrix}, \qquad \mathbf{D} = \begin{bmatrix} \mathbf{d}_L & \mathbf{0}_{15\times12} \\ \mathbf{0}_{15\times12} & \mathbf{d}_R \end{bmatrix}$$

#### B. Ellipsoid Constraint

Given that the state is estimated in the e-frame, we next propose a new Earth ellipsoid constraint to mitigate the height drift. Specifically, if one of the foot-mounted IMUs' height variation is less than a prescribed threshold $\varepsilon$ between two adjacent stance phases

$$\left| h\left( t_{k-m} \right) - h\left( t_k \right) \right| < \varepsilon \qquad (15)$$

where $t_{k-m}$ and $t_k$ are the times of previous and current stance phase, respectively, then it is reasonable to assume that the foot lies on a common ellipsoid surface during the neighboring stance phases, that is,

$$1 = \frac{\left( x^e \right)^2 + \left( y^e \right)^2}{\left( R_E + h \right)^2} + \frac{\left( z^e \right)^2}{\left( R_E \left( 1 - f^2 \right) + h \right)^2} + n_{ec} \qquad (16)$$

where $f$ denotes the earth eccentricity, $R_E$ is the transverse radius of curvature. $R_E$ and $h$ correspond to the state at $t_{k-m}$, and $n_{ec}$ denotes the ellipsoid constraint noise with covariance $\sigma_{ec}^2$.

Certainly, the two feet are both subject to their own ellipsoid constraints and the first-order linearized measurement equation is collectively given as

$$\begin{bmatrix} \delta y_{ec,L} \\ \delta y_{ec,R} \end{bmatrix} = \begin{bmatrix} \mathbf{H}_{ec,L} & \mathbf{0}_{1\times15} \\ \mathbf{0}_{1\times15} & \mathbf{H}_{ec,R} \end{bmatrix} \delta\mathbf{X} + \begin{bmatrix} n_{ec,L} \\ n_{ec,R} \end{bmatrix} \qquad (17)$$

where

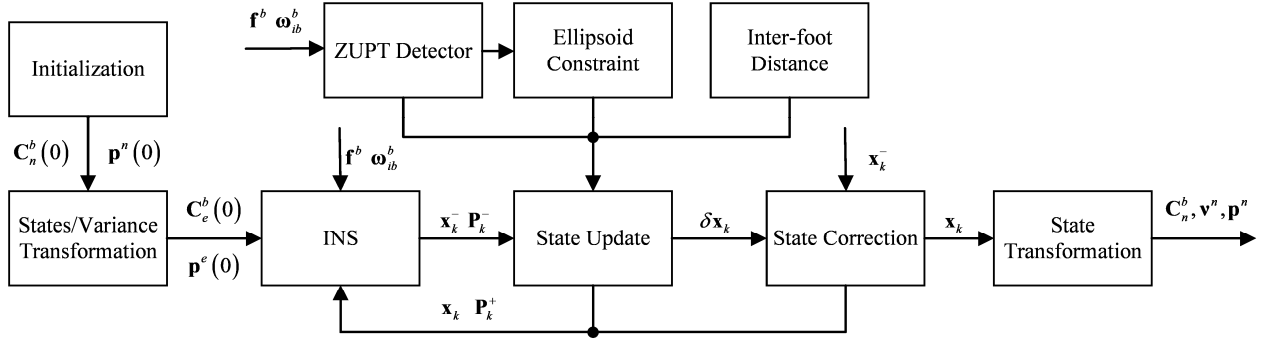

Fig. 2. Information flow in proposed pedestrian navigation system. Arrowed lines indicate information flow directions and associated symbols mean that their computation needs to feed on the source information.

$$\mathbf{H}_{ec} = \left[ \mathbf{0}_{1\times6} \quad \frac{2x^e}{\left(R_E + h\right)^2} \quad \frac{2y^e}{\left(R_E + h\right)^2} \quad \frac{2z^e}{\left(R_E\left(1-f^2\right)+h\right)^2} \quad \mathbf{0}_{1\times6} \right]$$
(18)

Figure 2 displays the information flow of the proposed PNS estimation algorithm. A user remains stationary for a few seconds to perform system initialization. The initial position is by a user input, the initial velocity and the accelerometer bias is assumed to be zero, the initial gyroscope bias is approximated by averaging the stationary gyroscope outputs and the initial attitude is obtained by accelerometer and magnetometer measurements [27]. Then the initial state and the corresponding covariance in the n-frame is converted to the e-frame [27]. Finally, the estimated state by the EKF is transformed to the n-frame.

### C. Observability Analysis

Many previous works have demonstrated that all states (except the position, the heading angle and the gyroscope bias about the gravity direction) are observable under the ZUPT measurement, see e.g. [8, 11]. Herein, we provide a global state observability analysis, proposed by the authors in [29, 30], of ZUPTs and inter-foot ranging, which highlights the observability benefit of the inter-foot ranging. The main idea of the global observability analysis is to constructively develop a sufficient condition for the initial state determination by making use of all inputs and outputs over the whole time interval under investigation. In contrast, the traditional observability methods based on the corresponding linearized system only use the local system information. Interested readers are referred to [29, 30] for details.

#### 1) Observability with ZUPT Measurement

It should be noted that the observability with ZUPT is performed in the n-frame so as to yield observability conclusion that is intuitive and for direction comparison with the previous literature.

*Theorem 1*: The level angles, gyroscope bias and the accelerometer bias are observable if and only if $\mathbf{K}$ has full column rank ($\mathbf{K}$ is defined in (29)).

*Proof*: Ignoring the Earth's rotation and measurement noise, the time derivative of body's velocity in n-frame is given as

$$\dot{\mathbf{v}}^n = \mathbf{C}_b^n\left(\mathbf{f}^b - \mathbf{b}_a\right) + \mathbf{g}^n$$
(19)

where $\mathbf{g}^n = \begin{bmatrix} 0 & -g & 0 \end{bmatrix}^T$. Assume the foot remains still at both the initial time zero and the time $t$. It means the velocity changes from time zero to $t$ is zero, i.e.,

$$\int_0^t \mathbf{C}_b^n\left(\mathbf{f}^b - \mathbf{b}_a\right) + \mathbf{g}^n dt = \mathbf{0}$$
(20)

By the chain rule of the attitude matrix and ignoring the changes of n-frame, $\mathbf{C}_b^n$ at time $t$ satisfies

$$\mathbf{C}_b^n\left(t\right) = \mathbf{C}_{n(t)}^{n(t)} = \mathbf{C}_{n(0)}^{n(t)}\mathbf{C}_{b(0)}^{n(0)}\mathbf{C}_{b(t)}^{b(0)} \approx \mathbf{C}_b^n\left(0\right)\mathbf{C}_{b(t)}^{b(0)}$$
(21)

where $\mathbf{C}_{b(t)}^{b(0)}$ denotes the attitude changes of the b-frame and $\mathbf{C}_b^n\left(0\right)$ is the initial attitude. Substituting (21) into (20) yields

$$\int_0^t \mathbf{C}_b^n\left(0\right)\mathbf{C}_{b(t)}^{b(0)}\left(\mathbf{f}^b - \mathbf{b}_a\right)dt + \mathbf{g}^n t = 0$$

$$\Rightarrow \int_0^t \mathbf{C}_{b(t)}^{b(0)}\left(\mathbf{f}^b - \mathbf{b}_a\right)dt + \begin{bmatrix} \sin\phi_{nb} \\ \cos\phi_{nb}\cos\theta_{nb} \\ -\sin\phi_{nb}\cos\theta_{nb} \end{bmatrix} g t = 0$$
(22)

where $\phi_{nb}$ and $\theta_{nb}$ are level angles. The yaw is eliminated above implies that it is unobservable by ZUPT. The integral term in Eq. (22) can be approximately calculated to the first order as (See details in Appendix B)

$$\int_0^t \mathbf{C}_{b(t)}^{b(0)}\left(\mathbf{f}^b - \mathbf{b}_a\right)dt \approx \chi\mathbf{b}_a + \gamma\mathbf{b}_g - \alpha$$
(23)

where $\chi$, $\gamma$ and $\alpha$ are expressed as

$$\alpha = -\sum_{k=0}^{M-1}\tilde{\mathbf{C}}_{b(t_k)}^{b(0)}\Delta\mathbf{v}$$

$$\chi = -\sum_{k=0}^{M-1}T\tilde{\mathbf{C}}_{b(t_k)}^{b(0)}\left[\mathbf{I}_3 + \frac{1}{6}\left(5\Delta\mathbf{\theta}_1 + \Delta\mathbf{\theta}_2\right)\times\right]$$

$$\gamma = \sum_{k=0}^{M-1}\left[T\sum_{i=1}^k\left(\tilde{\mathbf{C}}_{b(t_k)}^{b(0)}\Delta\mathbf{v}\times\tilde{\mathbf{C}}_{b(t_i)}^{b(0)}\mathbf{J}_r\left(\Delta\mathbf{\theta}\right)\right) + \frac{T}{6}\tilde{\mathbf{C}}_{b(t_k)}^{b(0)}\left(\Delta\mathbf{v}_1 + 5\Delta\mathbf{v}_2\right)\times\right]$$
(24)

where $\tilde{\mathbf{C}}_{b(t_k)}^{b(0)}$ denotes the erroneous rotation matrix from the initial time to $t_k$ computed by the error-contaminated body angle velocity $\mathbf{\omega}_{ib}^b$. $M$ is the number of samples in the time interval $[0,t]$, $\Delta\mathbf{\theta}_1$, $\Delta\mathbf{\theta}_2$ are the first and the second samples of the gyroscope-measured increment angle and $\Delta\mathbf{v}_1$, $\Delta\mathbf{v}_2$ are the first and the second samples of the accelerometer-measured incremental velocity, respectively, during the update interval $[t_k, t_{k+1}]$. The quantities $\Delta \mathbf{v}$, $\Delta \boldsymbol{\theta}$ and $\mathbf{J}_r$ are defined as

$$\Delta \boldsymbol{\theta} = \Delta \boldsymbol{\theta}_1 + \Delta \boldsymbol{\theta}_2$$
$$\Delta \mathbf{v} = \Delta \mathbf{v}_1 + \Delta \mathbf{v}_2 + \frac{1}{2}\Delta \boldsymbol{\theta} \times (\Delta \mathbf{v}_1 + \Delta \mathbf{v}_2) + \frac{2}{3}(\Delta \boldsymbol{\theta}_1 \times \Delta \mathbf{v}_2 + \Delta \mathbf{v}_1 \times \Delta \boldsymbol{\theta}_2)$$
(25)
$$\mathbf{J}_r(\Delta \boldsymbol{\theta}) = \frac{\sin\phi}{\phi}\mathbf{I}_3 + \left(1 - \frac{\sin\phi}{\phi}\right)\mathbf{aa}^T - \frac{1-\cos\phi}{\phi}\mathbf{a}\times \quad (26)$$

where $\phi \triangleq \|\Delta \boldsymbol{\theta}\|$, $\mathbf{a} = \Delta \boldsymbol{\theta}/\|\Delta \boldsymbol{\theta}\|$ and $\mathbf{J}_r$ is the right Jacobian of the special orthogonal whose determinant is +1, denoted $SO(3)$, as given in [31]. Therefore, Eq. (22) can be approximated to the first order as

$$\boldsymbol{\alpha} = \begin{bmatrix}\boldsymbol{\chi} & \boldsymbol{\gamma} & \boldsymbol{\eta}\end{bmatrix}\begin{bmatrix}\mathbf{b}_a \\ \mathbf{b}_g \\ \mathbf{x}_\theta\end{bmatrix} \triangleq \mathbf{kX} \quad (27)$$

where

$$\boldsymbol{\eta} = gMT\mathbf{I}_3$$
$$\mathbf{x}_\theta = \begin{bmatrix}\sin\phi_{nb} & \cos\phi_{nb}\cos\theta_{nb} & -\sin\phi_{nb}\cos\theta_{nb}\end{bmatrix}^T \quad (28)$$

For $n+1$ ZUPTs, $n$ constraints of the initial states can be similarly established. Let $\mathbf{y} = \begin{bmatrix}\boldsymbol{\alpha}_1^T & \cdots & \boldsymbol{\alpha}_n^T\end{bmatrix}^T$ and $\mathbf{K} = \begin{bmatrix}\mathbf{k}_1^T & \cdots & \mathbf{k}_n^T\end{bmatrix}^T$, where $\boldsymbol{\alpha}_n$ and $\mathbf{k}_n$ are the $\boldsymbol{\alpha}$ and $\mathbf{k}$ in the n-th constraints given in (27). The initial state $\mathbf{X}$ can be uniquely determined if and only if $\boldsymbol{\kappa}$ is a column full rank matrix.

$$\mathbf{X} = (\mathbf{K}^T\mathbf{K})^{-1}\mathbf{K}^T\mathbf{y} \quad (29)$$

Thus, the state $\mathbf{X}$ is observable if $\mathbf{K}$ is a column full matrix. ∎

The complexity of $\mathbf{K}$ does not allow a tractable rank analysis. Thus, we turn to a numerical investigation in the Section IV, which will show that $\mathbf{K}$ has full column rank if the foot-mounted IMU undergoes straight walking and a turning. Note that the requirement of straight walking and a turning is quite modest, that is to say, the requirement is almost certainly satisfied in practice. However, there exists a relatively small eigenvalue of $\mathbf{K}^T\mathbf{K}$ and the estimation of the gravity-direction gyroscope bias is not satisfying. These observations collectively lead us to conclude that the heading gyroscope bias is just weakly observable.

*2) Observability with Inter-foot Ranging*

The observability with the inter-foot ranging measurement is analyzed based on the premise of (two IMUs) known body's velocity, accelerometer bias, level angles and level gyroscope bias by ZUPT.

*Lemma 1* [30]: Given known points $\mathbf{a}_k$, $k = 1,2,...,m$, in three-dimensional space satisfying $\|\mathbf{a}_k - \mathbf{x}\| = r$, where $\mathbf{x}$ is an unknown point. If points $\mathbf{a}_k$ do not lie in any common plane, then $\mathbf{x}$ has a unique solution.

*Theorem 2*: For the case of small lever arm of the ultrasonic unit, the dual-IMUs' relative position in the e-frame $\mathbf{p}_{L,R}^e$ is observable.

*Proof*: Ignoring the measurement noise and the small lever arm, the distance measurement in (11) is given as

$$d(t) = \|\mathbf{p}_L^e(t) - \mathbf{p}_R^e(t)\| \quad (30)$$

where the feet position at time $t$ are calculated respectively by integrating (1)

$$\mathbf{p}_L^e(t) = \mathbf{p}_L^e(0) + \int_0^t \mathbf{v}_L^e d\tau$$
$$\mathbf{p}_R^e(t) = \mathbf{p}_R^e(0) + \int_0^t \mathbf{v}_R^e d\tau \quad (31)$$

Substituting the above equations into (30) yields

$$d(t) = \left\|\mathbf{p}_{L,R}^e(0) + \int_0^t (\mathbf{v}_L^e - \mathbf{v}_R^e)d\tau\right\| \quad (32)$$

where $\mathbf{p}_{L,R}^e(0) \triangleq \mathbf{p}_R^e(0) - \mathbf{p}_L^e(0)$ denotes the initial relative feet position expressed in the e-frame. Because the feet velocities are observable by the ZUPT measurement, Eq. (32) reveals that the solution of $\mathbf{p}_{L,R}^e(0)$ can be any point on the sphere surface with radius $d(t)$ and centering at $\int_0^t (\mathbf{v}_L^e - \mathbf{v}_R^e)d\tau$. According to *Lemma 1*, $\mathbf{p}_{L,R}^e(0)$ can be uniquely determined if $\int_0^t (\mathbf{v}_L^e - \mathbf{v}_R^e)d\tau$ for all times do not lie in any plane. The condition is quite moderate for normal walks that change directions now and then. Therefore, $\mathbf{p}_{L,R}^e$ is observable. ∎

*Theorem 3*: The dual-IMUs' relative heading is observable.
*Proof*: For both feet, we have with (1) and (2)

$$\dot{\mathbf{v}}_L^e = \mathbf{C}_L^e(\mathbf{f}_L^b - \mathbf{b}_{a,L}) - 2\boldsymbol{\omega}_{ie}^e \times \mathbf{v}_L^e + \mathbf{g}^e$$
$$\dot{\mathbf{v}}_R^e = \mathbf{C}_R^e(\mathbf{f}_R^b - \mathbf{b}_{a,R}) - 2\boldsymbol{\omega}_{ie}^e \times \mathbf{v}_R^e + \mathbf{g}^e \quad (33)$$

Attitude transformation does not change the magnitude of a vector, so subtracting the above two equations and taking the norm give

$$\left\|\dot{\mathbf{v}}_{L,R}^e + 2\boldsymbol{\omega}_{ie}^e \times \mathbf{v}_{L,R}^e\right\| = \left\|(\mathbf{f}_L^b - \mathbf{b}_{a,L}) - \mathbf{C}_L^R(\mathbf{f}_R^b - \mathbf{b}_{a,R})\right\| \quad (34)$$

where $\mathbf{v}_{L,R}^e = \mathbf{v}_R^e - \mathbf{v}_L^e$, $\dot{\mathbf{v}}_{L,R}^e = \dot{\mathbf{v}}_R^e - \dot{\mathbf{v}}_L^e$, $\mathbf{C}_L^R = \mathbf{C}_R^e\mathbf{C}_L^e$. Because the level angles decoded in $\mathbf{C}_L^R$ and the accelerometer bias are observable/known under the ZUPT measurement (according to Theorem 1), the heading angle is the only unknown variable in (34). Thus, the heading angle in $\mathbf{C}_L^R$ can be readily determined from the equality in (34). It means the dual-IMUs' relative heading is observable. ∎

*Theorem 4*: The dual-IMUs' relative gyroscope bias is observable.

*Proof*: Taking time derivative of $\mathbf{C}_L^R$ yields

$$\dot{\mathbf{C}}_L^R = \mathbf{C}_L^R\boldsymbol{\omega}_{RL}^L \times = \mathbf{C}_L^R\left[\boldsymbol{\omega}_{ib,L}^b - \mathbf{b}_{g,L} - \mathbf{C}_R^L(\boldsymbol{\omega}_{ib,R}^b - \mathbf{b}_{g,R})\right] \times \quad (35)$$

Define the relative gyroscope bias in the left foot frame as

$$\mathbf{b}_{L,R}^b \triangleq \mathbf{C}_R^L\mathbf{b}_{g,R} - \mathbf{b}_{g,L} \quad (36)$$

(35) can be expressed as

TABLE II
PARAMETER SETTING IN SIMULATION

| Stride length ( $l_s$ ) | 1.3m | Ranging noise ( $n_d$ ) | 0.02m |
|---|---|---|---|
| Max height ( $l_h$ ) | 0.14m | Max pitch ( $\theta_{\max}$ ) | 0.55rad |
| Swing time ( $t_u$ ) | 0.8s | Gyro noise ( $\mathbf{n}_b$ ) | 0.5 deg/$\sqrt{h}$ |
| Stance time ( $t_s$ ) | 0.4s | Acc. noise ( $\mathbf{n}_a$ ) | 0.001 m/s$^2$/$\sqrt{hz}$ |
| Turning time ( $t_a$ ) | 0.2s | | |
| Lever arm | $\mathbf{l}_L^b$=[0.02,0.05,-0.03] $\mathbf{l}_R^b$=[0.03, -0.03,0.04] | | |

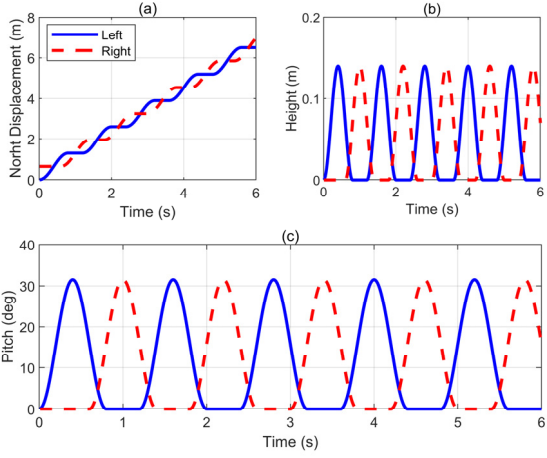

Fig. 3. Simulated states of a person walking from south to north in the first 6 seconds. (a): two feet displacement in north direction, (b): two feet displacement in up direction, (c): two feet pitch angle (solid blue line: left foot, dash green line: right foot)

$$\mathbf{b}_{L,R}^L \times = \mathbf{C}_R^L \boldsymbol{\omega}_{ib,R}^b \times \mathbf{C}_L^R - \boldsymbol{\omega}_{ib,L}^b \times + \mathbf{C}_R^L \dot{\mathbf{C}}_L^R \qquad (37)$$

As $\mathbf{C}_L^R$ is observable from Theorem 3, $\mathbf{b}_{L,R}^L$ can be uniquely determined, so is the relative gyroscope bias in the right foot frame by $\mathbf{b}_{L,R}^R = \mathbf{C}_L^R \mathbf{b}_{L,R}^L$ . Thus, the dual-IMUs' relative gyroscope bias is observable. ∎

As shown by the above analysis in Theorems 2-4, the inter-foot ranging can make the relative heading, position and gyroscope bias in the gravity direction observable. However, the absolute heading and heading gyroscope bias are still unobservable or weak observable. The Theorem below predicts that the full observability might be acquired by the inter-foot relative position vector. Note that the relative position measurement implies the relative ranging.

*Theorem 5*: If the dual-IMUs' relative position is measured, then the attitudes and the gyroscope biases of the dual-foot IMUs are totally observable.

*Proof*: The relationship between $\mathbf{p}_{L,R}^e$ and $\mathbf{p}_{L,R}^L$ is expressed as

$$\mathbf{p}_{L,R}^e = \mathbf{C}_{b,L}^e \mathbf{p}_{L,R}^L \qquad (38)$$

The level angles in $\mathbf{C}_{b,L}^e$ are known by ZUPT and the dual-IMUs' relative position in the e-frame, $\mathbf{p}_{L,R}^e$ is known by the inter-foot ranging according to Theorem 2. Thus, if the dual-IMUs' relative position in the left foot-IMU frame $\mathbf{p}_{L,R}^L$ is known by measurement, then the heading angle of the left-foot IMU can be solved by (38). Taking time derivative of $\mathbf{C}_{b,L}^e$ and ignoring the Earth rotation yield

$$\dot{\mathbf{C}}_{b,L}^e = \mathbf{C}_{b,L}^e \left( \boldsymbol{\omega}_{ib,L}^b - \mathbf{b}_{g,L} \right) \times \qquad (39)$$

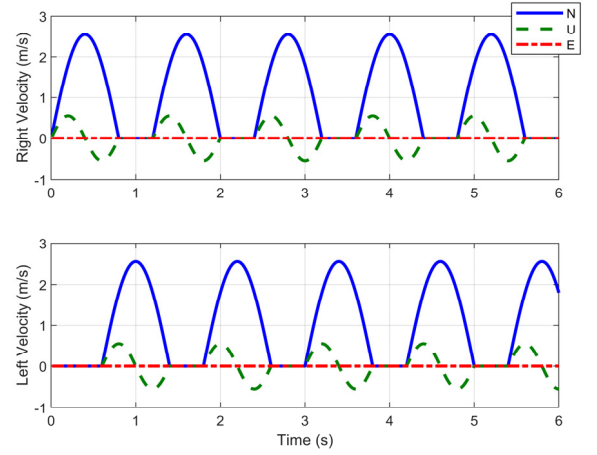

Fig. 4. Simulated velocity of a person waking from south to north in the first 6 seconds. (solid blue line: velocity in the north direction, dash green line: velocity in the up direction, dotted red line: velocity in the east direction.)

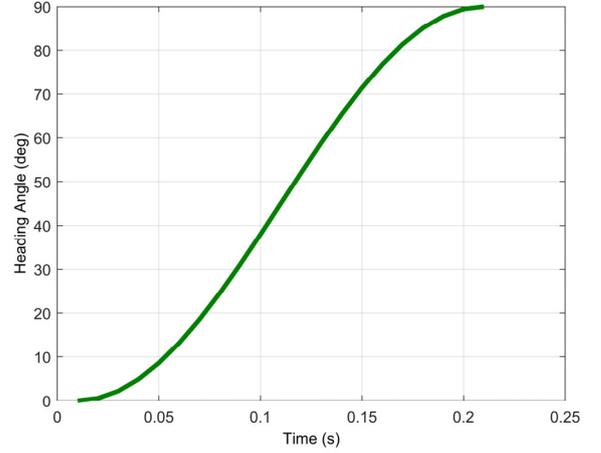

Fig. 5. Heading angle during a 90-degree right turn.

So the left-foot IMU's gyroscope bias can be calculated by (39)

$$\mathbf{b}_{g,L} \times = \boldsymbol{\omega}_{ib,L}^b \times - \mathbf{C}_{b,L}^e {}^T \dot{\mathbf{C}}_{b,L}^e \qquad (40)$$

Apparently, the above analysis also applies to the right foot. Hence, the attitudes and the gyroscope biases of the dual-foot IMUs are observable, if the dual-IMUs' relative position are available. ∎

Theorem 5 presents a picture of a promising improvement of the dual foot-mounted IMUs system. And, the inter-foot relative position might be measured by using a camera on one foot to observe the known points on the other foot.

## IV. SIMULATION RESULTS

In this section, the performance of ZUPT and ZUPT+RNG are compared by simulating a person walking around a square with two feet mounted IMU and inter foot ranging. The

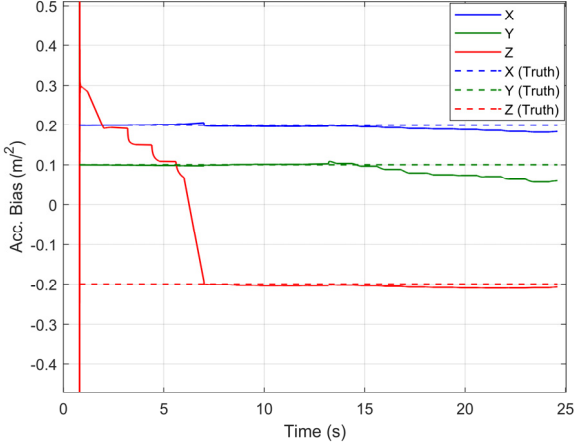

Fig. 6. Accelerometer bias estimate by least square in (29) (solid line: estimated value, dash line: truth value; blue line: accelerometer bias in X-axis direction, green line: accelerometer bias in Y-axis direction, red line: accelerometer bias in Z-axis direction)

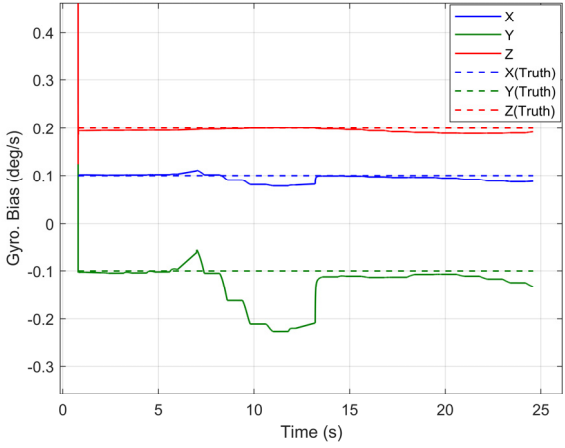

Fig. 7. Gyroscope bias estimate by least square in (29) (solid line: estimated value, dash line: truth value; blue line: gyroscope bias in X-axis direction, green line: gyroscope bias in Y-axis direction, red line: gyroscope bias in Z-axis direction)

observability of ZUPT as well as ZUPT+RNG analyzed in Section III are also verified by simulation.

### A. Simulation Settings

Assume one walks along a straight line in the horizontal plane with the heading angle $\psi_0$. The initial position, velocity and Euler angle of the left foot in the n-frame are $\mathbf{p}_L^n(0) = \begin{bmatrix} \lambda_0 & L_0 & h_0 \end{bmatrix}^T$, $\mathbf{v}_L^n(0) = \begin{bmatrix} 0 & 0 & 0 \end{bmatrix}^T$ and $\boldsymbol{\theta}_{n,L}^b = \begin{bmatrix} 0 & \psi_0 & 0 \end{bmatrix}^T$, respectively.

For simplicity, each walking step is divided into two phases: swing phase and stance phase (with time durations $t_u$ and $t_s$) and the period of one step is $T = t_u + t_s$. In the $k$th ($k \geq 0$) swing phase, the relative position of the left foot with respect to the last stance phase $\Delta\mathbf{p} = \begin{bmatrix} p_n & p_u & p_e \end{bmatrix}^T$ is set to

$$p_n = l_s \left\{ 1 - \cos\left[ \pi(t - kT)/t_u \right] \right\} \cos(\psi_0)/2$$
$$p_u = l_h \left\{ 1 - \cos\left[ 2\pi(t - kT)/t_u \right] \right\}/2 \qquad (41)$$
$$p_e = l_s \left\{ 1 - \cos\left[ \pi(t - kT)/t_u \right] \right\} \sin(\psi_0)/2$$

where $p_n$, $p_u$ and $p_e$ are relative displacements in north, up and east directions, respectively. $l_s$ denotes the length of one stride and $l_h$ denotes the maximum vertical displacement in each step. The absolute position of the left foot is calculated by

$$\mathbf{p}_L^n(t) = f\left( \Delta\mathbf{p}, \mathbf{p}_L^n(kT) \right) \qquad (42)$$

where $f(\bullet)$ is a function to calculate the position in the n-frame according to a known position in the n-frame and the relative position (See detail in Appendix A). The velocity of the left foot $\mathbf{v}_L^n = \begin{bmatrix} v_n & v_u & v_e \end{bmatrix}^T$ can be calculated by taking time derivative of (41)

$$v_n = l_s \pi \sin\left[ \pi(t - kT)/t_u \right] \cos(\psi_0)/2t_u$$
$$v_u = l_h \pi \sin\left[ 2\pi(t - kT)/t_u \right]/t_u \qquad (43)$$
$$v_e = l_s \pi \sin\left[ \pi(t - kT)/t_u \right] \sin(\psi_0)/2t_u$$

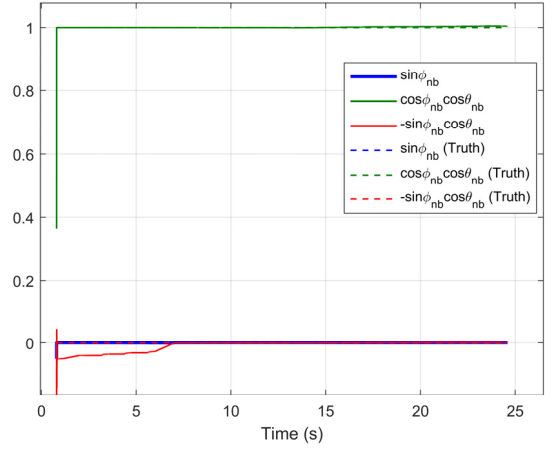

Fig. 8. Level angles estimate by least square in (29) (solid line: estimated value, dash line: truth value;)

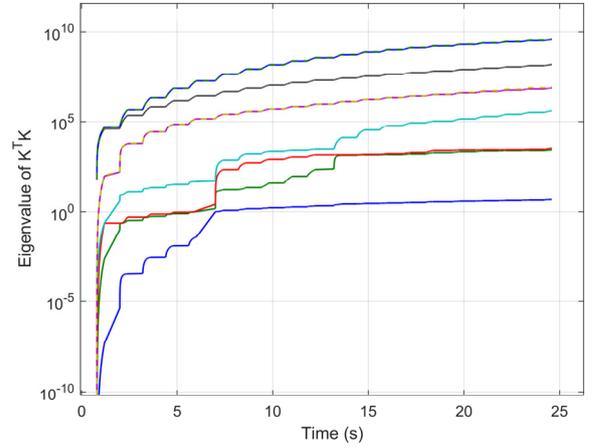

Fig. 9. Eigenvalue of $\mathbf{K}^T\mathbf{K}$ in (29)

| | Estimated Position (m) | Position Error (m) | Relative Position Error (m) | Yaw Error (deg) | Relative Yaw Error (deg) | Yaw Bias Error (deg/s) | Relative Yaw Bias Error (deg/s) |
|---|---|---|---|---|---|---|---|
| Left/Right (ZUPT) | (41.71,10.72)/(-4.34 25.91) | 43.05/26.42 | 48.58 | 121.25/71.16 | 167.6 | 0.132/0.075 | 0.207 |
| Left/Right (ZUPT + RNG) | (-0.17,0.22)/(0.50, 0.85) | 0.28/0.25 | 0.027 | 2.22/2.28 | 0.06 | 0.0022/0.0005 | 0.0017 |
| Truth | Position (Left/Right): (0,0)/(0.65,0.65) m | | | Yaw: 0 deg | | Gyro Bias: (2,2.3,1.7) deg/s | |

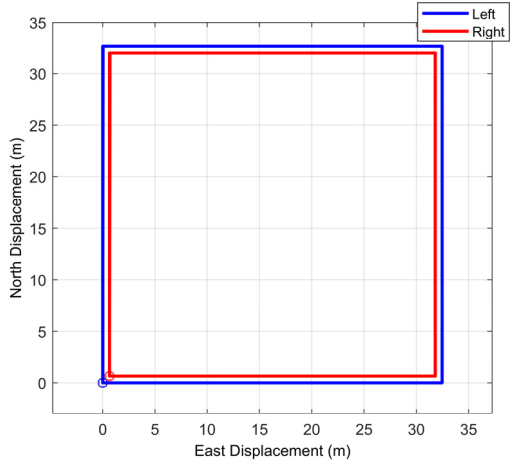

Fig. 10. Simulated trajectory in horizontal plane (circle: start point; square: end point; blue line: left foot, red line: right foot.)

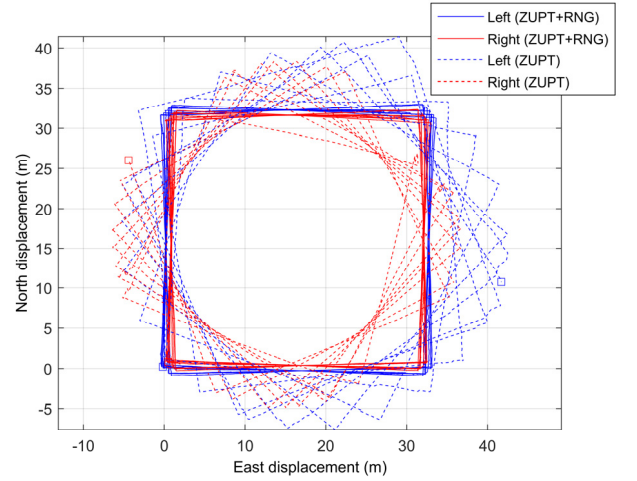

Fig. 11. Estimated trajectory in the horizontal plane (circle: start point; square: end point; solid line: ZUPT+RNG, dash line: ZUPT; blue line: left foot, red line: right foot. )

And the attitude of the left foot $\boldsymbol{\theta}_{n,L}^{b} = \begin{bmatrix} \phi_{nb} & \psi_{nb} & \theta_{nb} \end{bmatrix}^{T}$ is expressed as

$$\phi_{nb} = 0 \quad \psi_{nb} = \psi_{0}$$
$$\theta_{nb} = \theta_{\max}\left\{1 - \cos\left[2\pi\left(t - kT\right)/t_{u}\right]\right\}/2 \quad (44)$$

where $\theta_{\max}$ is the maximum pitch during walking. As for the stance phase, the position, velocity and attitude are equal to those at the end time of the last swing phase.

In order to get a close-loop trajectory, we also simulate a person making turns of 90 degrees with zero position change. If the heading at the start of the turning is $\psi_{0}$, then the attitude of the left foot is given by

$$\phi_{nb} = 0 \quad \theta_{nb} = 0$$
$$\psi_{nb} = \pi\left\{1 - \cos\left[\pi\left(t - t_{o}\right)/t_{r}\right]\right\}/4 + \psi_{0} \quad (45)$$

where $t_{r}$ is the duration of turning and $t_{o}$ denotes the start time of turning. The position during the turning remains unchanged and the velocity is set to zero.

The simulated left-foot gyroscope output is given as

$$\boldsymbol{\omega}_{ib}^{b} = \boldsymbol{\omega}_{nb}^{b} + \mathbf{C}_{e}^{b}\mathbf{C}_{e}^{n}\boldsymbol{\omega}_{ie}^{e} + \mathbf{b}_{g} + \mathbf{n}_{g} \quad (46)$$

where $\boldsymbol{\omega}_{nb}^{b}$ denotes the angular rate of the left-foot IMU with respect to the n-frame. Because there is only one Euler angle changing at a time, $\boldsymbol{\omega}_{nb}^{b}$ can be calculated by

$$\boldsymbol{\omega}_{nb}^{b} = \dot{\boldsymbol{\theta}}_{n,L}^{b} \quad (47)$$

and the simulated left-foot accelerometer output is given as

$$\mathbf{f}^{b} \approx \mathbf{C}_{n}^{b}\left(\dot{\mathbf{v}}^{n} - \mathbf{g}^{n} + 2\boldsymbol{\omega}_{ie}^{n} \times \mathbf{v}^{n}\right) + \mathbf{b}_{a} + \mathbf{n}_{a} \quad (48)$$

The sampling rate of IMU is 100Hz. The right-foot IMU output is the same as the left-foot IMU, except that the right foot is delayed by $T/2$ in time to simulate the alternate swing of the two feet. In addition, the initial position of the right foot in the $n_{0}$-frame is

$$\mathbf{p}_{R}^{n_{0}} = \left[\frac{l_{s}}{2}\left(\cos\psi_{0} + \sin\psi_{0}\right) \quad 0 \quad \frac{l_{s}}{2}\left(\cos\psi_{0} - \sin\psi_{0}\right)\right]^{T} \quad (49)$$

where the superscript $n_{0}$ denotes the local tangent plane frame with the origin at the initial position of the left foot (x-axis: north, y-axis: up, z-axis: east). Therefore, (49) means that the right foot is half stride to the right and half stride forward of the left foot. Then, the initial position of right foot in the n-frame $\mathbf{p}_{R}^{n}$ can be obtained by

$$\mathbf{p}_{R}^{n}\left(0\right) = f\left(\mathbf{p}_{R}^{n_{0}}, \mathbf{p}_{R}^{n}\left(0\right)\right) \quad (50)$$

Finally, the inter-foot distance is given by (11) at 10 Hz. Table II lists the parameter setting in our simulation. The trajectory profile of a person walking from south to north in the first 6 seconds is shown in Figs. 3-4. Figure 3 shows two feet displacements in the north and height directions as well as the pitch variation. The displacement in the east direction, the heading and roll angles are not given, since these values remain unchanged. The walking speed is plotted in Fig. 4. In the swing phase, the velocities in the north and height directions are simulated by using a trigonometric function. The heading angle during a 90-degree turn is shown in Fig. 5 and other states are unchanged during the turning process.

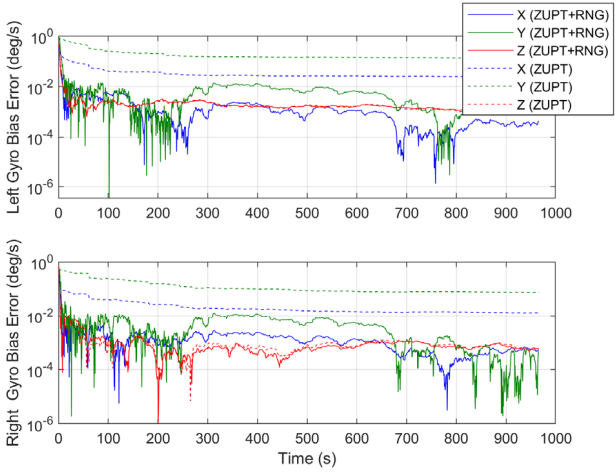

Fig. 12. Estimate error of gyroscope bias (solid line: ZUPT+RNG, dash line: ZUPT; blue line: gyroscope bias error in X-axis direction, green line: gyroscope bias error in Y-axis direction, red line: gyroscope bias error in Z-axis direction)

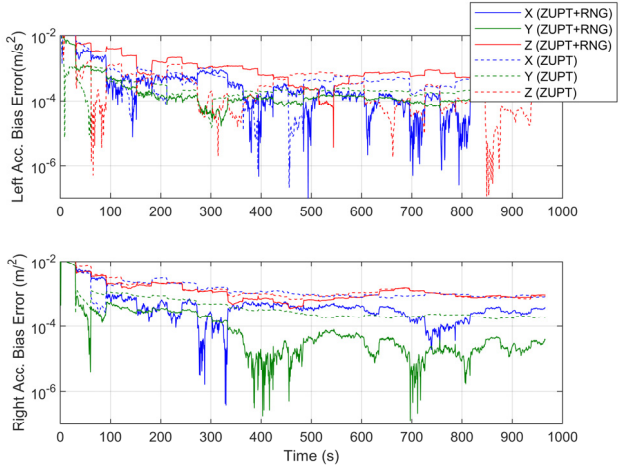

Fig. 13. Estimate error of accelerometer bias. (solid line: ZUPT+RNG, dash line: ZUPT; blue line: accelerometer bias error in X-axis direction, green line: accelerometer bias error in Y-axis direction, red line: accelerometer bias error in Z-axis direction)

*B. Estimation Results*

First, the ZUPT observability (in Theorem 1) is demonstrated by a simulated square trajectory lasting 24 seconds (taking right turns at 6s, 12s and 18s), for which the inter-foot ranging measurement is absent. The IMU's true accelerometer and gyroscope bias are set to $\begin{bmatrix} 0.2 & 0.1 & -0.2 \end{bmatrix}$ m/s$^2$ and $\begin{bmatrix} 0.05 & -0.05 & 0.06 \end{bmatrix}$ deg/s, respectively, and the three initial attitudes are all zero degrees. Figures 6-8 plot the estimated accelerometer and gyroscope biases as well as the level angles obtained by (29). We see that all states, except the z-axis accelerometer bias (in Fig. 6) and a state related to level angles (in Fig. 8), converge to the corresponding truths from the very first swing at 0.8s. After the first turning at 6s, the two states

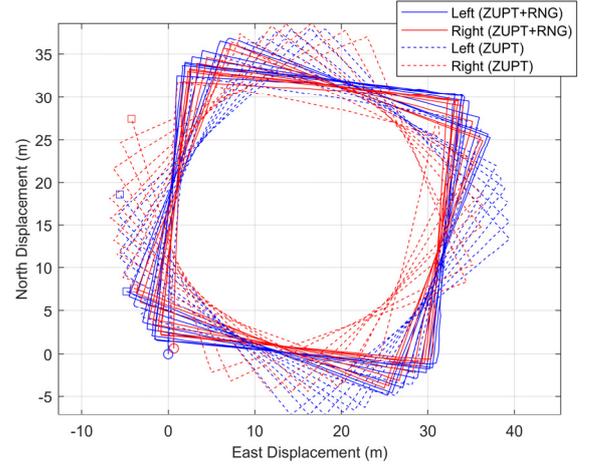

Fig. 14. Estimated trajectory in the horizontal plane (circle: start point; square: end point; solid line: ZUPT+RNG, dash line: ZUPT; blue line: left foot, red line: right foot.)

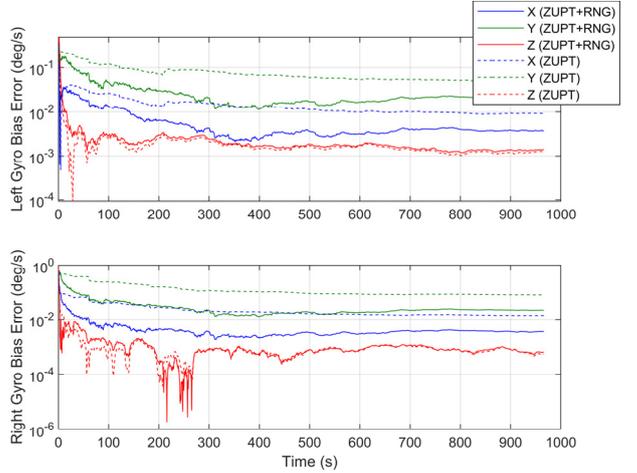

Fig. 15. Estimate error of gyroscope bias (solid line: ZUPT+RNG, dash line: ZUPT; blue line: gyroscope bias error in X-axis direction, green line: gyroscope bias error in Y-axis direction, red line: gyroscope bias error in Z-axis direction)

converge as well. Compared with other states, the gyroscope bias in y-axis (gravity direction) is estimated not so well, which implies that the y-axis gyroscope is weakly observable. This is consistent with the eigenvalues of $\mathbf{K}^T\mathbf{K}$, as shown in Fig. 9. All eigenvalues become nonzero after the first turning but there is an eigenvalue that stays relatively quite small throughout the simulation. This means that one state in $\mathbf{X}$ is nearly unobservable or weakly observable. Note that the deviation of the estimates from their truths, as apparent in Figs. 6-7, is owed to the approximation error in formulating (27) that inevitably accumulates along with time, but the estimation results are a convincing support to the global observability analysis of ZUPT in Theorem 1. The weak observability of the gravity-direction gyroscope bias is a new insight over the established results in the literature.

Next, we turn to evaluate the performance of the proposed PNS algorithm compared with the ZUPT only method. Similarly, we simulate a person walking around a square (side length $25l_s$) for eight circles and stopping where he starts. The initial heading is set to zero and the person takes a 90-degree right turn at every corner, and the total travelled distance is 1040 meters with 966.4 seconds. For both IMUs, the true gyroscope bias is $\begin{bmatrix} 2 & 2.3 & 1.7 \end{bmatrix}$ deg/s and the true accelerometer bias is $\begin{bmatrix} 0.1 & 0.2 & -0.2 \end{bmatrix}$ m/s². Figure 10 plots the simulated trajectory in the horizontal plane.

TABLE IV
Start-End Estimated Error for Two Feet

| | Estimated Position (m) | Position Error (m) | Relative Position Error (m) | Yaw Error (deg) | Relative Yaw Error (deg) | Yaw Bias Error (deg/s) | Relative Yaw Bias Error (deg/s) |
|---|---|---|---|---|---|---|---|
| Left/Right (ZUPT) | (-5.60,18.65)/ (-4.26 27.45) | 19.47/27.25 | 8.18 | 53.98/74.25 | 20.27 | 0.050/0.080 | 0.03 |
| Left/Right (ZUPT + RNG) | (-4.82,7.27)/ (-3.98, 7.62) | 8.72/8.37 | 0.36 | 22.85/22.92 | 0.07 | 0.020/0.022 | 0.002 |
| Truth | Position (Left/ Right): (0,0)/(0.65,0.65) m | | | Yaw: 0 deg | | Gyro Bias: (2,2.3,1.7) deg/s | |

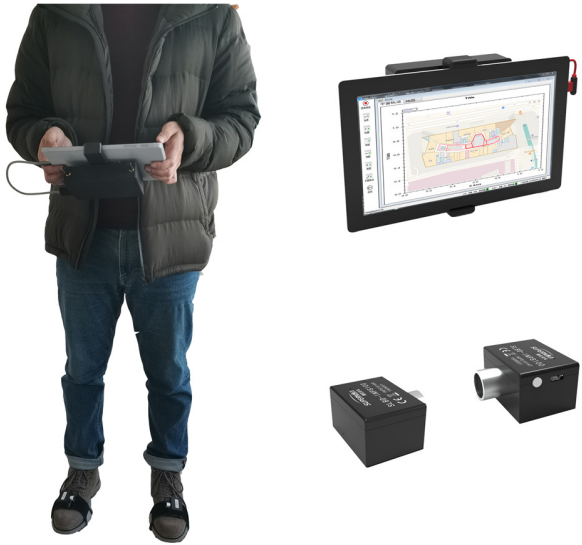

Fig. 16. Left: System overview. Upper right: data processing/display module. Low right: IMU and ultrasonic ranging modules

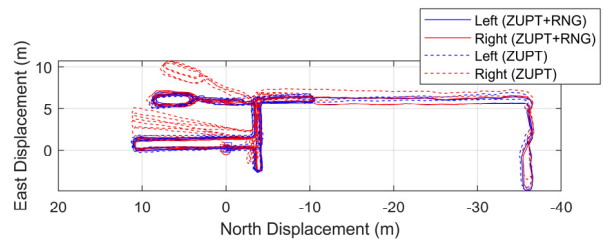

Fig. 17. Estimated trajectory in horizontal plane (circle: start point; square: end point; solid line: ZUPT+RNG, dash line: ZUPT; blue line: left foot, red line: right foot.).

The initial position of the left foot is set to $\mathbf{p}_n^v(0) = \begin{bmatrix} 121E & 31N & 0 \end{bmatrix}$, and the initial right-foot position is computed by (50). The initial attitude errors for the left-foot and the right-foot IMUs are $\boldsymbol{\theta}_{n,L}^b = \begin{bmatrix} 2° & 5° & 2° \end{bmatrix}^T$, $\boldsymbol{\theta}_{n,R}^b = \begin{bmatrix} -2° & -3° & -4° \end{bmatrix}^T$, respectively. The initial velocity and the accelerometer bias are set to zero. The initial gyroscope bias of the left-foot IMU is set to $\mathbf{b}_{g,L} = \begin{bmatrix} 1.7 & 1.6 & 1.3 \end{bmatrix}^T$ deg/s, while that of the right-foot IMU is set to $\mathbf{b}_{g,R} = \begin{bmatrix} 2.5 & 2.8 & 1 \end{bmatrix}^T$ deg/s (the left/right heading gyro bias error is set to -0.7/+0.5 deg/s, respectively. The IMU noise statistics in the filter is set to be equal to the true noise statistics and the measurement variances for ZUPT and inter-foot ranging are set to $0.05^2 \mathbf{I}_3$ and $0.05^2$, respectively.

Figures 11-13 plot the estimated results by the ZUPT method and the proposed method of ZUPT plus inter-foot ranging. As shown by the observability analysis in Section III, the accelerometer bias and the x-axis, z-axis (horizontal plane) gyroscope bias are correctly estimated by both methods. However, due to the uncorrected initial heading error and the weakly-observable gyroscope bias in y-axis (gravity direction), the heading estimation by the ZUPT method diverges gradually, as evidenced in Fig. 11. While, with the aid of inter-foot ranging, the heading drift is largely eliminated. The start-end dual feet' absolute and relative estimation errors are listed in Tables III. Although the inter-foot ranging suppresses the heading and position error dramatically, the absolute error still remains. Table III clearly shows that the dual-IMUs' relative heading, position and heading bias are largely corrected, which is consistent with the observability analysis in Section III.

In order to show that the absolute heading gyro bias is weakly observable, we further simulate the same trajectory as above, but the initial gyroscope bias of the left/right-foot IMUs is set to $\mathbf{b}_{g,L} = \begin{bmatrix} 1.7 & 2.6 & 1.3 \end{bmatrix}^T$ deg/s and $\mathbf{b}_{g,R} = \begin{bmatrix} 2.5 & 2.8 & 1 \end{bmatrix}^T$ deg/s (the left/right heading gyro bias error is set to +0.3/+0.5 deg/s, respectively). The estimated trajectory and the gyroscope bias are displayed in Figs 14-15. Although the ZUPT-RNG method performs better than the ZUPT-only method does, the heading angle and the heading gyro bias in the proposed method have not yet been fully corrected, as shown in Table IV.

## V. Experiments

### A. Hardware Implementation

We designed and implemented a real-time PNS system with dual foot-mounted IMUs and ultrasonic ranging, as shown in Fig. 16. The system consists of two parts: sensor modules and data processing/display module. The sensor module is composed of an Xsens MTi-1 MEMS IMU chip with $10°/h$ gyro bias stability and a ST202/CX20106A ultrasonic transmitter and receiver chip, while the data processing/display module includes a data receiver and a tablet computer. The data acquisition process is described as follows: the wireless synchronizing trigger signal is first transmitted by the sensor module, then the time-synchronized IMU and ultrasonic data collected via an STM32 single chip is sent to data processing/display module by a 2.4G wireless transmission link.

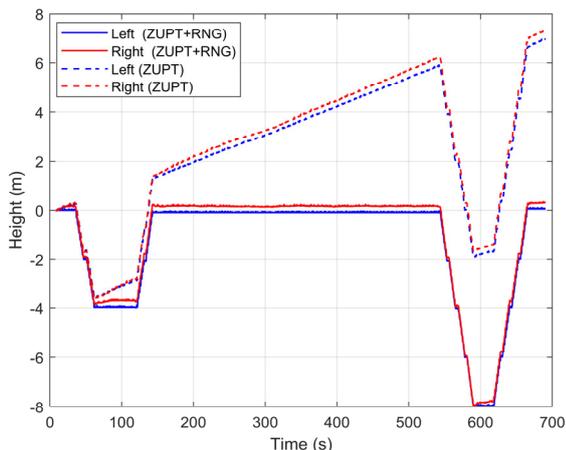

Fig. 18. Height estimate. (solid line: ZUPT+RNG, dash line: ZUPT; blue line: left foot, red line: right foot.)

To validate the ultrasonic ranging accuracy, the measured inter-foot ranging is calibrated by a Vicon system [32]. The calibration results show that the ranging accuracy is about 3 cm (RMS).

### B. Test Results

This subsection is devoted to verifying the proposed PNS algorithms by both indoor and outdoor experiments. In the indoor experiment, the tester walks along a corridor for about 30 seconds at the very beginning, then goes downstairs and upstairs by one floor, circles the corridor several times, walks downstairs and upstairs by two floors and finally returns to the starting point. The total time duration is about 700 seconds with walking distance of about 431m. Figure 17 displays the estimated trajectory in the horizontal plane, which shows that ZUPT cannot well compensate the heading angle, especially for the right foot. However, with the aid of inter-foot ranging, the heading error is significantly mitigated. As for the height estimate, unlike the overestimated height by the ZUPT-only method, the proposed PNS algorithm yields a quite satisfying height estimate, as shown in Fig. 18. Figure 19 further displays the 3D trajectory and Table V summarizes the start-end error for both methods.

In the first outdoor experiment, one tester walks along a square and the other tester follows right behind him with a Samsung S10 mobile phone held in hand. The outdoor experiment lasts about 700 seconds and the total travelled distance is about 632m. Figure 20 compares the GPS positioning result and the estimated trajectory by averaging the feet position by the proposed PNS algorithm. Due to a tall building nearby the walking path that incurs signal blockage or multipath, the GPS positioning accuracy is just about 5m, while the start-end error of the proposed PNS reaches about 0.7m (0.11% of the travelled distance). A longer test is conducted with travelled distance about 1600m and total time 28 min. One tester starts from a gym and walks along the campus paths. After taking a few turns, the tester goes around a round square twice, then crosses a lake, and finally comes back to the round square. The test results of both methods are shown in Fig. 21. Comparing with the road in Google Map, the max error of the proposed method is about 8 m (0.5 % of travelled distance), and the end error is less than 3 m (0.18% of travelled distance).

| Position Error | Left/Right (ZUPT) | Left/Right (ZUPT +RNG) |
|---|---|---|
| Level Plane (m) | 0.55/2.25 | 0.14/0.31 |
| Percentage of Travelled Distance (%) | 0.13/0.5 | 0.03/0.07 |
| Height (m) | 6.96/7.31 | 0.09/0.31 |

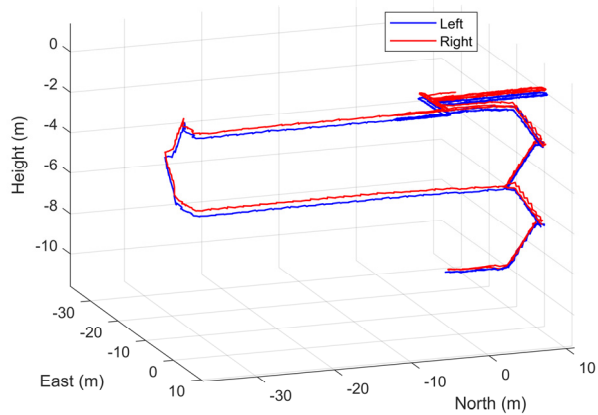

Fig. 19. Estimated 3D trajectory by ZUPT + RNG PNS algorithm. (blue line: left foot, red line: right foot.)

## VI. CONCLUSIONS AND FUTURE WORKS

In this paper, we have developed a pedestrian navigation system equipped with dual foot-mounted low-cost IMUs and foot-to-foot ranging. The global observability analysis shows that the inter-foot ranging, together with the ZUPT, can effectively mitigate the heading error that has been an inherent problem in any PNS. In specific, the relative heading, gyroscope bias and position between the dual IMUs are made observable by the inter-foot ranging. A new ellipsoid constraint is proposed to reduce the height drift. Simulations are designed to verify the observability analysis and demonstrate the PNS capability. Test results show that the proposed PNS algorithm is promising to deliver a positioning accuracy of 0.1-0.2% traveled distance for GNSS-denied or infrastructure-independent applications.

The global observability analysis also indicates a potential improvement of the implemented system by incorporating a relative vectoral position measurement of the feet.

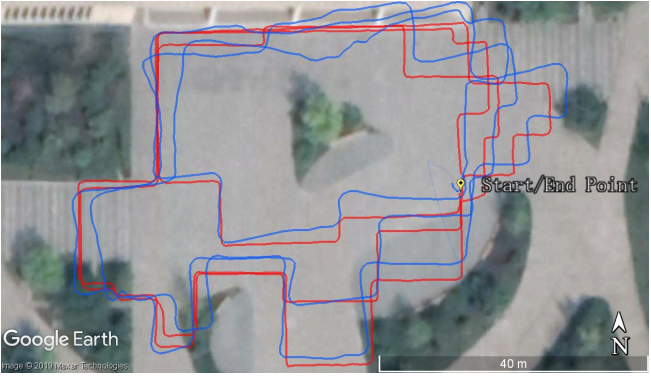

Fig. 20. Outdoor walking trajectory beside tall buildings overplayed with Google Earth (blue line: Samsung S10 GPS positioning; red line: ZUPT + RNG PNS positioning result).

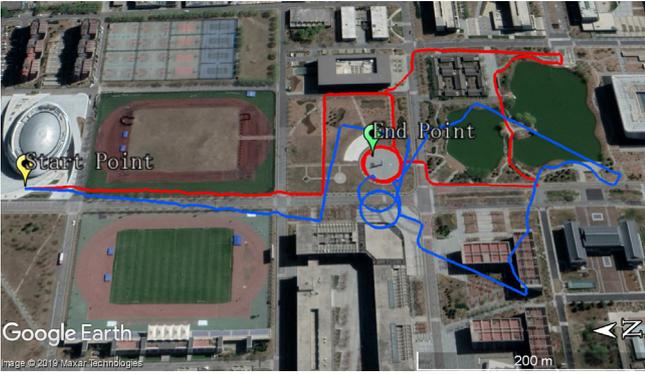

Fig. 21. Estimated trajectory of longer outdoor walking (blue line: ZUPT positioning result for left foot; red line: ZUPT + RNG positioning result)

## APPENDIX A

The function $\mathbf{p}_R^n = f\left(\Delta\mathbf{p}, \mathbf{p}_L^n\right)$ in (42) and (50) transforms the position in the tangent plane frame to the $n$ frame:

Step 1: Transform $\Delta\mathbf{p}$ into the relative position in the e-frame by

$$\Delta\mathbf{p}^e = \begin{bmatrix} \cos\lambda & 0 & -\sin\lambda \\ -\sin L\sin\lambda & \cos L & -\sin L\cos\lambda \\ \cos L\sin\lambda & \sin L & \cos L\cos\lambda \end{bmatrix}\Delta\mathbf{p} \quad (51)$$

Step 2: Transform $\mathbf{p}_L^n$ to $\mathbf{p}_L^e$ and calculate $\mathbf{p}_R^e$ by

$$\mathbf{p}_R^e = \Delta\mathbf{p}^e + \mathbf{p}_L^e \quad (52)$$

Step 3: Transform $\mathbf{p}_R^e$ to $\mathbf{p}_R^n$.

## APPENDIX B

### INTEGRAL APPROXIMATION IN (23)

The integral term in the right side of (23) was approximated by our group in [33]. Herein, we present an alternative and more accurate first-order approximation of this integral on manifold similar to [34].

The exponential map between rotation matrix and rotation vector and their first-order approximation are [31]

$$\mathbf{C} = \exp\left(\boldsymbol{\varphi}\times\right) \approx \mathbf{I}_3 + \boldsymbol{\varphi}\times \quad (53)$$

where $\mathbf{C} \in SO(3)$ and $\boldsymbol{\varphi}\times \in so(3)$.

For a small rotation vector $\delta\boldsymbol{\varphi}$, there exist the following properties [31]

$$\exp\left[\left(\boldsymbol{\varphi} + \delta\boldsymbol{\varphi}\right)\times\right] \approx \exp\left(\boldsymbol{\varphi}\times\right)\exp\left[\left(\mathbf{J}_r\left(\boldsymbol{\varphi}\right)\delta\boldsymbol{\varphi}\right)\times\right] \quad (54)$$

and

$$\exp\left(\boldsymbol{\varphi}\times\right)\exp\left(\delta\boldsymbol{\varphi}\times\right) \approx \exp\left[\left(\boldsymbol{\varphi} + \mathbf{J}_r\left(\boldsymbol{\varphi}\right)^{-1}\delta\boldsymbol{\varphi}\right)\times\right] \quad (55)$$

where $\mathbf{J}_r$ is the right Jacobian of $SO(3)$ given in (26).

Another useful property of exponential map is [31]

$$\exp\left(\boldsymbol{\varphi}\times\right)\mathbf{C} = \mathbf{C}\exp\left[\left(\mathbf{C}^T\boldsymbol{\varphi}\right)\times\right] \quad (56)$$

Depending on Eqs. (53)-(56), the matrix $\mathbf{C}_{b(t)}^{b(0)}$ in the right side of (23) can be approximated to the first order as

$$\begin{aligned} \mathbf{C}_{b(t)}^{b(0)} &= \mathbf{C}_{b(T)}^{b(0)}\cdots\mathbf{C}_{b(MT)}^{b((M-1)T)} \overset{(53)}{\approx} \prod_{i=1}^{M}\exp\left[\left(\Delta\boldsymbol{\theta}_i - T\mathbf{b}_g\right)\times\right] \overset{(55)}{\approx} \prod_{i=1}^{M}\exp\left(\Delta\boldsymbol{\theta}_i\times\right)\exp\left[-\left(\mathbf{J}_r\left(\Delta\boldsymbol{\theta}_i\right)T\mathbf{b}_g\right)\times\right] \\ &= \prod_{i=1}^{M-2}\exp\left(\Delta\boldsymbol{\theta}_i\times\right)\exp\left[-\left(\mathbf{J}_r\left(\Delta\boldsymbol{\theta}_i\right)T\mathbf{b}_g\right)\times\right]\exp\left(\Delta\boldsymbol{\theta}_{M-1}\times\right)\exp\left[-\left(\mathbf{J}_r\left(\Delta\boldsymbol{\theta}_{M-1}\right)T\mathbf{b}_g\right)\times\right]\exp\left(\Delta\boldsymbol{\theta}_M\times\right)\exp\left[-\left(\mathbf{J}_r\left(\Delta\boldsymbol{\theta}_M\right)T\mathbf{b}_g\right)\times\right] \\ &\overset{(56)}{\underset{(53)}{=}} \prod_{i=1}^{M-2}\exp\left(\Delta\boldsymbol{\theta}_i\times\right)\exp\left[-\left(\mathbf{J}_r\left(\Delta\boldsymbol{\theta}_i\right)T\mathbf{b}_g\right)\times\right]\tilde{\mathbf{C}}_{b(t)}^{b((M-2)T)}\exp\left[-\left(\tilde{\mathbf{C}}_{b(t)}^{b((M-1)T)T}\mathbf{J}_r\left(\Delta\boldsymbol{\theta}_{M-1}\right)T\mathbf{b}_g\right)\times\right]\exp\left[-\left(\mathbf{J}_r\left(\Delta\boldsymbol{\theta}_M\right)T\mathbf{b}_g\right)\times\right] \\ &\overset{(56)}{=} \tilde{\mathbf{C}}_{b(t)}^{b(0)}\prod_{i=1}^{M}\exp\left[-\left(\tilde{\mathbf{C}}_{b(t)}^{b(iT)T}\mathbf{J}_r\left(\Delta\boldsymbol{\theta}_i\right)T\mathbf{b}_g\right)\times\right] \end{aligned} \quad (57)$$

Let $\boldsymbol{\zeta}_i \triangleq -\tilde{\mathbf{C}}_{b(t)}^{b(iT)T}\mathbf{J}_r\left(\Delta\boldsymbol{\theta}_i\right)T\mathbf{b}_g$, (57) can be further reduced to

$$\begin{aligned} \mathbf{C}_{b(t)}^{b(0)} &\approx \tilde{\mathbf{C}}_{b(t)}^{b(0)}\prod_{i=1}^{M}\exp\left(\boldsymbol{\zeta}_i\times\right) \overset{(53)}{\approx} \tilde{\mathbf{C}}_{b(t)}^{b(0)}\prod_{i=1}^{M}\left(\mathbf{I}_3 + \boldsymbol{\zeta}_i\times\right) \\ &\approx \tilde{\mathbf{C}}_{b(t)}^{b(0)}\left(\mathbf{I}_3 + \sum_{i=1}^{M}\boldsymbol{\zeta}_i\times\right) \\ &= \tilde{\mathbf{C}}_{b(t)}^{b(0)}\left[\mathbf{I}_3 - \sum_{i=1}^{M}\left(\tilde{\mathbf{C}}_{b(t)}^{b(iT)T}\mathbf{J}_r\left(\Delta\boldsymbol{\theta}_i\right)T\mathbf{b}_g\right)\times\right] \end{aligned} \quad (58)$$

where the second-order and above term about $\mathbf{b}_g$ are ignored.

Assume the body angular rate and the specific force to be approximated in linear forms as

$$\begin{aligned} \boldsymbol{\omega}_{ib}^b &= t\boldsymbol{\alpha}_w + \mathbf{b}_w \\ \mathbf{f}^b &= t\boldsymbol{\alpha}_f + \mathbf{b}_f \end{aligned} \quad (59)$$

where $\boldsymbol{\alpha}_w$, $\mathbf{b}_w$, $\boldsymbol{\alpha}_f$, $\mathbf{b}_f$ are the coefficient vectors. The incremental angles are represented as

$$\Delta\boldsymbol{\theta}_1 = \int_0^{T/2} \boldsymbol{\omega}_{ib}^b dt = \int_0^{T/2} t\boldsymbol{\alpha}_w + \mathbf{b}_w dt = \frac{T^2}{8}\boldsymbol{\alpha}_w + \frac{T}{2}\mathbf{b}_w$$

$$\Delta\boldsymbol{\theta}_1 + \Delta\boldsymbol{\theta}_2 = \int_0^T \boldsymbol{\omega}_{ib}^b dt = \int_0^T t\boldsymbol{\alpha}_w + \mathbf{b}_w dt = \frac{T^2}{2}\boldsymbol{\alpha}_w + T\mathbf{b}_w \qquad (60)$$

from which the coefficient vectors are solved as

$$T^2\boldsymbol{\alpha}_w = 4\left(\Delta\boldsymbol{\theta}_2 - \Delta\boldsymbol{\theta}_1\right)$$
$$T\mathbf{b}_w = 3\Delta\boldsymbol{\theta}_1 - \Delta\boldsymbol{\theta}_2 \qquad (61)$$

Similarly, the coefficients of incremental velocities are

$$T^2\boldsymbol{\alpha}_f = 4\left(\Delta\mathbf{v}_2 - \Delta\mathbf{v}_1\right)$$
$$T\mathbf{b}_f = 3\Delta\mathbf{v}_1 - \Delta\mathbf{v}_2 \qquad (62)$$

Now we turn to the derivation of Eq. (23)

$$\int_0^t \mathbf{C}_{b(t)}^{b(0)}\left(\mathbf{f}^b - \mathbf{b}_a\right)dt \approx \sum_{k=0}^{M-1}\int_{t_k}^{t_{k+1}} \mathbf{C}_{b(t)}^{b(0)}\left(\mathbf{f}^b - \mathbf{b}_a\right)dt$$
$$= \sum_{k=0}^{M-1}\mathbf{C}_{b(t_k)}^{b(0)}\int_{t_k}^{t_{k+1}} \mathbf{C}_{b(t)}^{b(t_k)}\left(\mathbf{f}^b - \mathbf{b}_a\right)dt$$
$$\overset{(58)}{\approx} \sum_{k=0}^{M-1}\tilde{\mathbf{C}}_{b(t_k)}^{b(0)}\left[\mathbf{I}_3 - \sum_{i=1}^k\left(\tilde{\mathbf{C}}_{b(t_k)}^{b(t_i)\,T}\mathbf{J}_r\left(\Delta\boldsymbol{\theta}_i\right)T\mathbf{b}_g\right)\times\right]$$
$$\cdot\int_{t_k}^{t_{k+1}}\left(\mathbf{I}_3 + \left(\int_{t_k}^t \boldsymbol{\omega}_{ib}^b - \mathbf{b}_g d\tau\right)\times\right)\left(\mathbf{f}^b - \mathbf{b}_a\right)dt \qquad (63)$$

where the above incremental integral can be approximated by the two-sample correction using (59) and (62)

$$\int_{t_k}^{t_{k+1}}\left(\mathbf{I}_3 + \left(\int_{t_k}^t \boldsymbol{\omega}_{ib}^b - \mathbf{b}_g d\tau\right)\times\right)\left(\mathbf{f}^b - \mathbf{b}_a\right)dt$$
$$= \int_{t_k}^{t_{k+1}}\left(\mathbf{I}_3 + \left(\int_{t_k}^t \boldsymbol{\omega}_{ib}^b d\tau\right)\times\right)\mathbf{f}^b dt - \int_{t_k}^{t_{k+1}}\left(\mathbf{I}_3 + \left(\int_{t_k}^t \boldsymbol{\omega}_{ib}^b d\tau\right)\times\right)dt\mathbf{b}_a$$
$$- \mathbf{b}_g \times \int_{t_k}^{t_{k+1}}\left(t - t_k\right)\left(\mathbf{f}^b - \mathbf{b}_a\right)dt$$
$$= \Delta\mathbf{v} - T\left[\mathbf{I} + \frac{1}{6}\left(5\Delta\boldsymbol{\theta}_1 + \Delta\boldsymbol{\theta}_2\right)\times\right]\mathbf{b}_a$$
$$+ \left[\frac{T}{6}\left(\Delta\mathbf{v}_1 + 5\Delta\mathbf{v}_2\right) - \frac{T^2}{2}\mathbf{b}_a\right]\times\mathbf{b}_g \qquad (64)$$

where $\Delta\mathbf{v}$ is defined in (25). Note that there was a coefficient error of $\mathbf{b}_g$ in [33]. Substituting (64) into (63) yields (23).

ACKNOWLEDGMENT

Gratitude to Prof. Jade Morton in University of Colorado for constructive comments on this paper and thanks to graduate Mr. Luyang Zhou who contributed to an early prototype of the presented PNS system.